\documentclass[twocolumn]{aastex7} 

\usepackage{caption}
\usepackage{subcaption}
\usepackage{multirow}
\usepackage{mathtools,amssymb}
\usepackage{booktabs}
\usepackage{hyperref}
\usepackage{scrextend}

\newcommand{\age}{t_{\rm age}}
\newcommand{\xiinj}{\xi_{\rm inj}}
\newcommand{\vsh}{v_{\rm sh}}
\newcommand{\nism}{n_{\rm ISM}}
\newcommand{\ncl}{n_{\rm cl}}
\newcommand{\mpr}{m_{\rm p}}

\newcommand{\pcr}{P_{\rm CR}}

\newcommand{\w}[1]{v_{\rm A,#1}}

\newcommand{\rsh}{R_{\rm sh}}
\newcommand{\Kep}{K_{\rm ep}}
\newcommand{\Eej}{E_{\rm SN}}
\newcommand{\Mej}{M_{\rm ej}}
\newcommand{\f}{\textit{f}}
\newcommand{\Rcl}{R_{\rm cl}}
\newcommand{\Btwo}{B_{2}}
\newcommand{\xicr}{\xi_{\rm CR}}

\shorttitle{SN 1006}

\begin{document}

\title{SN 1006: A Cosmic Laboratory for Investigating Shock Acceleration Physics}

\correspondingauthor{Emma McGinness}
\email{emcginness@uchicago.edu}

\author[0000-0003-1457-2213]{Emma McGinness}
\email{emcginness@uchicago.edu}
\affiliation{Department of Physics, The University of Chicago, 5720 S Ellis Ave, Chicago, IL 60637, USA}

\author[0000-0002-6679-0012]{Rebecca Diesing}
\email{}
\affiliation{School of Natural Sciences, Institute for Advanced Study, Princeton, NJ 08540, USA}
\affiliation{Department of Physics and Columbia Astrophysics Laboratory, Columbia University, New York, NY 10027, USA}

\author[0000-0003-0939-8775]{Damiano Caprioli}
\email{}
\affiliation{Department of Astronomy and Astrophysics, The University of Chicago, 5640 S Ellis Ave, Chicago, IL 60637, USA}
\affiliation{Enrico Fermi Institute, The University of Chicago, Chicago, IL 60637, USA}

\author[0000-0002-6606-2816]{Fabio Acero}
\email{}
\affiliation{Université Paris-Saclay, Université Paris Cité, CEA, CNRS, AIM, F-91191, Gif-sur-Yvette Cedex, France}
\affiliation{FSLAC IRL 2009, CNRS/IAC, E-38205, La Laguna, Tenerife, Spain}

\begin{abstract}

SN 1006 is a historical Type Ia supernova remnant that exhibits nonthermal emission ranging from radio to multi-TeV $\gamma$-rays. Most of this emission (particularly X-rays and $\gamma$-rays) is concentrated in polar caps 
aligned with the ambient magnetic field, which makes it an ideal laboratory for studying cosmic ray (CR) acceleration at different shock obliquities and the hadronic/leptonic nature of the $\gamma$-ray emission. 
We model SN 1006's morphology, multiwavelength spectrum, and radial profile using a self-consistent multizone kinetic model of particle acceleration that accounts for CR-driven shock modification, magnetic field amplification, drift in magnetic fluctuations, and temporal dynamics including adiabatic and synchrotron losses.  
Our model can reproduce both the observed spectral and spatial properties, with the exception of the radio profile, which we argue requires 3D hydrodynamic effects to replicate.
We find that quasi-parallel regions (where the shock normal aligns with the ambient magnetic field) exhibit very prominent CR acceleration ($20\%$ efficiency), while quasi-perpendicular regions exhibit efficiencies below 1\%, consistent with the results of kinetic simulations. 
We also find that electrons are responsible for the majority of the $\gamma$-ray emission from SN 1006 (i.e., it is a leptonic source), with the exception of the northwest region due to an encounter with a dense cloud.  \\ 

\end{abstract}


\section{Introduction} \label{sec:intro}

Supernova remnants (SNRs) are thought to be sources of Galactic cosmic rays (CRs), possibly up to PeV energies, because their energetics and particle acceleration mechanism are well suited to reproduce the Galactic CR flux \citep[e.g.,][]{hillas05, berezhko+07, ptuskin+10, caprioli+10a}.  
Further, abundant observational evidence, such as the detection of synchrotron and hadronic $\gamma$-ray emission, supports SNRs as effective particle accelerators \citep[e.g.,][]{morlino+12, ackermann+13, slane+14,lemoine-goumard+22,humensky+25}.

In the standard paradigm, CR acceleration in SNRs occurs at the forward shock via diffusive shock acceleration (DSA), whereby magnetic perturbations scatter charged particles repeatedly across the shock front and particles gain energy with each crossing \citep{krymskii77,bell78a, blandford+78}. 
For relativistic particles, DSA yields power-law energy distributions $dN/dE \propto E^{-q}$, where $q = (R+2)/(R-1)$ is related to the fluid density compression ratio $R$.
For a strong shock with Mach number $M \gg 1$, $R\simeq4$ and, accordingly, $q=2$.
However, if we take into account nonlinear modifications to DSA associated with efficient CR acceleration, such as CR-driven magnetic field amplification (MFA) and CR escape, $R$ and $q$ deviate from standard DSA, allowing $R\geq4$.
While for many years nonlinear modifications were thought to yield CR spectra \textit{flatter} than $E^{-2}$ \citep[e.g.,][and references within]{kang+06,amato+06,caprioli+09a,caprioli+08}, accounting for the movement of the CR scattering centers with respect to the background plasma may actually yield spectra \textit{steeper} than $E^{-2}$ \citep{zirakashvili+08b,caprioli12}. 
Recently, kinetic simulations \citep{haggerty+20,caprioli+20} demonstrated that this is indeed the case, and in particular that efficient CR acceleration can provide both $R>4$ \citep[as inferred in Tycho and SN 1006 SNRs,][]{warren+05, gamil+08, giuffrida+22} and $q>2$ \citep[][and references therein]{caprioli11, bell+11, diesing+21}.

Nonthermal broadband emission from radio to $\gamma$-rays reveals the presence of accelerated particles.
However, the identity of particles producing $\gamma$-rays can be ambiguous: $\gamma$-rays may come from both hadronic processes, in which neutral pions produced in proton-proton interactions subsequently decay, and leptonic processes, in which relativistic electrons upscatter background photons via inverse Compton (IC) or emit relativistic bremsstrahlung.
In general, the nature of the $\gamma$-ray emission may depend on environmental parameters such as ambient density and local infrared/optical photon backgrounds \citep{KBV05, ellison+07, morlino+09, corso+23}.

\begin{figure}
    \centering
    \includegraphics[width=0.45\textwidth]{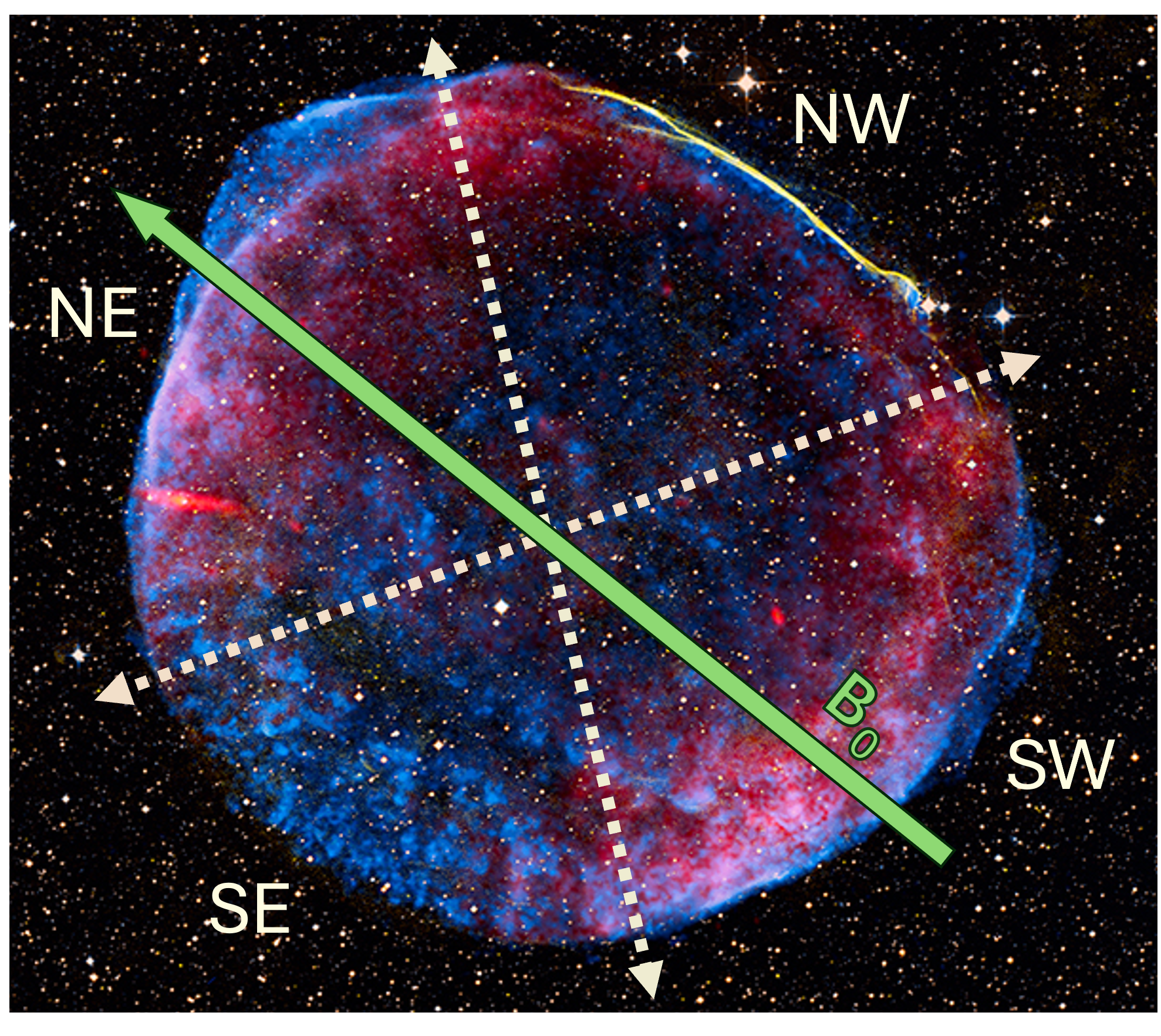}
    \caption{\label{fig:quad} SN 1006 composite image in X-ray (blue), radio (red), and optical (yellow/orange) bands.  The quadrants are sectioned (dashed arrows) and labeled (white).  The ambient magnetic field $B_0$ (green arrow) is quasi-parallel to the NE/SW limbs and quasi-perpendicular to the NW/SE limbs.  The NW limb is radially compressed.  Image credit: X-ray: NASA/CXC/Rutgers/G.Cassam-Chena\"{\i}, J.Hughes et al.; radio: NRAO/AUI/NSF/GBT/VLA/Dyer, Maddalena $\&$ Cornwell; optical: Middlebury College/F.Winkler, NOAO/AURA/NSF/CTIO Schmidt $\&$ DSS.}
\end{figure}

Here, we study particle acceleration in SN 1006, which is generally accepted to be the remnant of a Type Ia SN explosion based on location (high above the Galactic plane), ejecta chemical composition, and historical light curve \citep[see][]{winkler+03}.
SN 1006 is roughly spherical, with bilateral morphology; nonthermal X-ray and TeV emission are predominantly confined within two polar caps, as shown in Figure \ref{fig:quad} \citep{koyama+95, allen+01, SN1006HESS}.
Additionally, radio fractional polarization measurements \citep{reynoso+13}\footnote{\label{fn:fracpol}
A reliable radio fractional polarization measurement of SN 1006 is unattainable from recent interferometric observations because its angular size exceeds the maximum scale recoverable by MeerKAT \citep{cotton+24,zhou+25}; hence we rely on maps from VLA and ATCA for this quantity \citep{reynoso+13}. Nevertheless, the observations are consistent in downstream field orientation along the polar caps.} reveal that the downstream magnetic field is disordered in the polar caps, and ordered elsewhere.
The orientation of the ambient magnetic field, determined from radio and X-ray morphology \citep[e.g.,][]{rothenflug+04, bocchino+11} and radio polarization \citep[e.g.,][]{reynoso+13, cotton+24}, suggests a relationship between this azimuthal variation and shock obliquity; regions where the shock normal is quasi-parallel to the ambient magnetic field (i.e., the \textit{polar caps}) exhibit efficient CR acceleration and strong MFA, and conversely quasi-perpendicular regions (i.e., nonpolar caps) exhibit comparatively inefficient particle acceleration, negligible MFA, and minimal nonthermal emission at high energies \citep[][and references therein]{giuffrida+22, lemoine-goumard+25}.
This is consistent with the findings of kinetic shock simulations, which show that spontaneous injection of ions into DSA and efficient acceleration occur preferentially for quasi-parallel configurations \citep[e.g.,][]{sironi+09, guo+13, caprioli+14a, caprioli+15}.
Moreover, the frequency cutoff of the nonthermal X-ray synchrotron emission exhibits an azimuthal variation along the polar caps, which suggests a dependence of CR acceleration and transport on shock obliquity \citep[][and references within]{gamil+08,katsuda+09,miceli+16}.

In this paper, we perform a spatially resolved analysis of SN 1006 to test 
the theoretical prediction that quasi-parallel shocks are more efficient than oblique ones in accelerating particles to high energies (up to multi-TeV) and how the ambient density impacts ``hadronicity" \citep[a term we use to refer to the ratio of hadronic to leptonic $\gamma$-ray emission as defined in][]{corso+23}.

SN 1006 is an ideal cosmic laboratory because (1) it has been extensively observed with good spatial resolution at multiple wavelengths \citep[e.g.,][and citations within]{allen+01,bamba+08,SN1006HESS,lemoine-goumard+25}; (2) it is embedded in an almost uniform ambient magnetic field (implied by its bilateral morphology); and (3) it is expanding into a density gradient approximately perpendicular to the field axis (determined from postshock thermal X-ray emission, e.g., \citealp{long+03,acero+07,katsuda+13,giuffrida+22} and Balmer line emission, e.g., \citealp{ghavamian+02,heng+07,raymond+07,bandiera+19}).

SN 1006 can be sectioned into four quadrants based on shock obliquity and ambient density profile (Figure \ref{fig:quad}), where the northeast (NE) and southwest (SW) quadrants comprise the polar caps.
The ambient density of SN 1006 is roughly uniform \citep[$\lesssim0.05$ $\text{cm}^{-3}$, e.g.,][]{acero+07,miceli+12,winkler+14,giuffrida+22} except in the northwest (NW), where there is evidence of recent shock interaction with a denser atomic cloud \citep[e.g.,][and citations within]{acero+07,raymond+07,winkler+13,katsuda+13,bandiera+19}.
Several papers have modeled the morphology and the spectral energy distribution (SED) of a single quadrant \citep[e.g.,][and citations therein]{allen+08,morlino+10,miceli+16,giuffrida+22} and others have considered multiquadrant spectral \citep[e.g.,][]{acero+07,tao+24,lemoine-goumard+25} and spatial \citep[e.g.,][]{miceli+09, winner+20} properties.

What differentiates this paper is that we account for the multiwavelength spectral and spatial information concurrently across the entire SNR using a self-consistent kinetic particle acceleration model based on the solution of the diffusion--advection equation for the transport of CRs \citep[see][and references therein]{caprioli12, diesing+21}, which contains state-of-the-art nonlinear DSA theory arising from kinetic plasma simulations. 
By combining all available spatial and spectral information in addition to requiring consistency with kinetic theory, we reduce the number of free parameters to the point that our model is overconstrained.

We outline our method in \S\ref{sec:model}, show our results alongside observations and discuss our findings in \S\ref{sec:results}, before concluding  in \S\ref{sec:conclusion}.



\section{Model} \label{sec:model}

We describe our framework for modeling the nonthermal emission of each quadrant in SN 1006.  Our model assumes all nonthermal emission originates from the FS, which is typically the case for Type Ia SNRs that expand in a homogeneous medium \citep[e.g.,][ which ruled out any contribution from the reverse shock in Tycho]{morlino+12}.

\subsection{Shock Hydrodynamics} \label{subsec:hydro}

We consider the evolution of a typical Type Ia SNR with ejecta energy $\Eej = 10^{51}$ ergs and ejecta mass $\Mej = 1.2$ $\text{M}_{\odot}$ in one dimension, assuming spherical symmetry.   
For the polar caps and SE region, we assume uniform ambient density $n_0(r) = \nism$.  For the NW region, we model $n_0(r)$ as a step function to account for the presence of a dense atomic cloud with density $\ncl$ at a distance $\Rcl$ from the remnant's center.  
A discontinuous density profile could explain the bright Balmer filament \citep{bandiera+19} and radial compression (see Figure \ref{fig:quad}) observed in the NW.

We take the ambient magnetic field to be $B_0 = 3$ $\mu$G, and the ambient temperature $T_{\rm ISM} = 10^4$ K.
Because SN 1006 is a relatively young SNR ($\sim 1020$ yr old), we consider two stages of SNR evolution: the ejecta-dominated (ED) stage, where $\Mej$ is greater than the swept-up mass, and the Sedov-Taylor (ST) stage, where the shock expands adiabatically.  

For the ED stage, we use the analytical solution of \citet{tang+17}, with a power-law ejecta structure of index $n=7$ and shock expansion into a homogeneous medium ($s=0$). 
For our reference $\nism$, SN 1006 has only recently exited the ED stage (within the last $\sim200$ yr) with the exception of the NW region, where the shock encounters a dense cloud that immediately drives the transition from the ED to ST stage, considerably earlier than elsewhere; this occurs when $\rsh = \Rcl$ with $\vsh = v_{\rm j}$.

We determine the shock radius where the ST stage begins ($R_{\rm ST}$), and corresponding shock velocity ($v_{\rm ST}$), using the condition that the transition from ED to ST occurs when $\Mej$ equals the swept-up mass, and solving for $R_{\rm ST}$ analytically.

During the ST stage, we use the thin-shell approximation, where the SNR mass is taken to be contained within a thin shell behind the shock front \citep{bisnovatyi-kogan+95, ostriker+88, bandiera+04}, which allows us to accommodate nonuniform density profiles. 
$\vsh$ is calculated using the solution in \cite{ptuskin+05} for a thin shell with mass \mbox{$M(\rsh) = 2\Mej + \int_{R_\star}^{\rsh} 4 \pi r^2 \mpr\, n_0(r)\,dr$}, i.e.,
\begin{equation}
\begin{split}
\vsh &= \Bigg[ \left(\frac{R_\star}{\rsh}\right)^{\omega}
      \frac{4\Mej^2}{M^2(\rsh)}\, v_\star^2 \\
   &\quad + \frac{\omega \Eej (\gamma_{\rm ad}+1)^2}{2 \rsh^{\omega} M^2(\rsh)}
      \int_{R_\star}^{\rsh} r^{\omega-1} M(r)\,dr \Bigg]^{1/2} ,
\end{split}
\label{eq:vsh}
\end{equation}
where \mbox{$\omega = 6(\gamma_{\rm ad}-1)/(\gamma_{\rm ad}+1)$} with $\gamma_{\rm ad}=5/3$ the adiabatic index for a monatomic gas, and $(R_\star,v_\star)$ is $(\Rcl,v_{\rm j})$ for the NW quadrant and $(R_{\rm ST},v_{\rm ST})$ elsewhere.
Shock radius and velocity are given as a function of time using $dt = dr/\vsh$.

We use observations constraining SN 1006 properties to inform our model.  
We take the age of SN 1006 to be $\age = 1020$ yr \citep{stephenson10} and the distance to be 2.2 kpc \citep{winkler+03,giuffrida+24}.
We estimate the observed outer shock radius (for NE, SW, and SE regions) to be $\sim\!9.1$ pc from the observed angular distances in \citet{gamil+08}, \citet{SN1006HESS}, and \citet{katsuda+13}.
For the NW region, we approximate its radius by drawing two concentric circles on Figure \ref{fig:quad}, with the larger radius proportional to the outer shock radius, and use the ratio of their radii to get $\sim 7.6$ pc.
We set $\Rcl = 5$ pc based on results from \cite{acero+07}, which concluded that the shock collided with a dense cloud recently, and the NW radius of $\sim 7.6$ pc.

\begin{figure}
    \begin{centering}
    \includegraphics[width=0.48\textwidth]{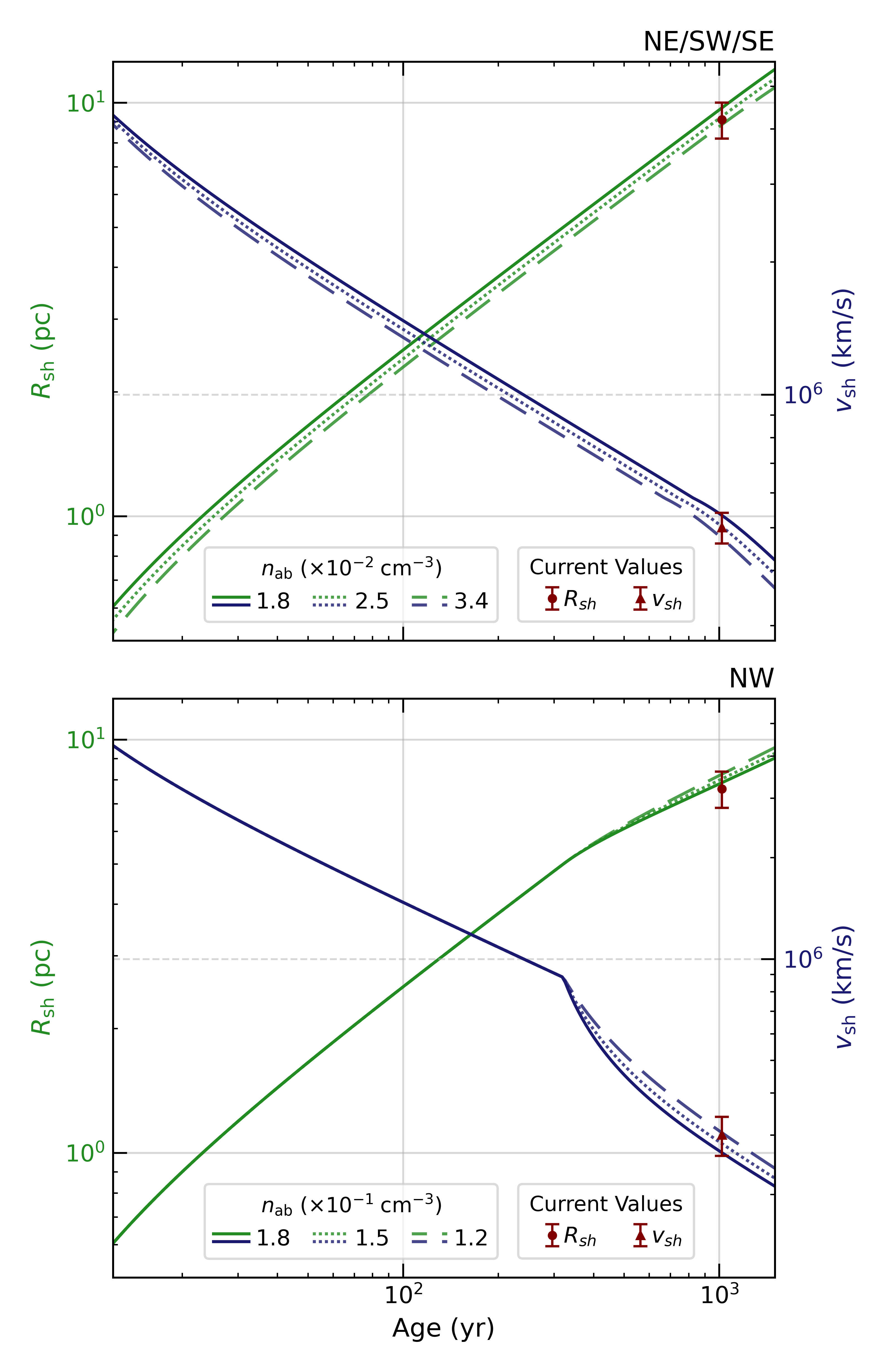}
    \caption{\label{fig:evol} Evolution of shock radius and velocity (green and blue curves/axes, respectively) for the quadrants identified in the title, assuming $\Eej = 10^{51}$ ergs and $\Mej = 1.2$ $\text{M}_{\odot}$. 
    All curves are consistent with the measurements of $v_{\rm sh,age} = 5000 \pm 400$ $\rm km\,s^{-1}$ and $R_{\rm sh,age} = 9.1 \pm 0.9$ pc for the top panel, and $v_{\rm sh,age} = 3000 \pm 400$ $\rm km\,s^{-1}$ and $R_{\rm sh,age} = 7.6 \pm 0.8$ pc for the bottom panel.
    Different line styles correspond to different ambient density $\nism\sim0.018-0.034$ cm$^{-3}$ (top panel) and cloud density $0.12-0.18$ cm$^{-3}$ (bottom panel).
    The sudden change in slope at $\sim 600$ yr is attributed to the density jump from $\nism = 1.8\times10^{-2}$ cm$^{-3}$ to $\ncl$ at $\Rcl = 5$ pc.}
    \end{centering}
\end{figure}

The shock velocities estimated from proper motions are $5000 \pm 400$ $\rm km\,s^{-1}$ for the NE, SW, and SE regions \citep{katsuda+09} and $3000 \pm 400$ $\rm km\,s^{-1}$ for the NW region \citep{ghavamian+02,winkler+03}.
These constrain $\nism$ to $0.018-0.034$ cm$^{-3}$ (Figure \ref{fig:evol}, top panel) and $\ncl$ to $0.12-0.18$ cm$^{-3}$ (Figure \ref{fig:evol}, bottom panel).
We favor $\nism = 1.8 \times10^{-2}$ $\text{cm}^{-3}$ based on $\vsh$ at $\age$ ($v_{\rm sh,age}$) and the observed hadronic emission of the polar caps \citep{lemoine-goumard+25}, a value in reasonable agreement with X-ray measurements \citep{acero+07,miceli+12,winkler+14,giuffrida+22}, too.
Through comparison with Fermi-LAT observations in the NW \citep{lemoine-goumard+25}, we set $\ncl = 0.18$ $\text{cm}^{-3}$, roughly consistent with \citet{ghavamian+02}, \citet{long+03}, \citet{heng+07}, \citet{raymond+07}, and \citet{bandiera+19}. 
With these parameters, both $v_{\rm sh,age}$ and $R_{\rm sh,age}$ (i.e., $\rsh$ at $\age$) result within $10\%$ of the inferred values (see Figure \ref{fig:evol}).  
This holds regardless of whether we assume $\age$ is the present age (1020~yr) or the age when the radio and X-ray maps were taken  \citep[$\sim1010$~yr;][]{gamil+08,SN1006HESS,katsuda+13}.

\subsection{Particle Acceleration}  \label{subsec:paracc}

We use a semianalytic multizone kinetic model of nonlinear DSA, as in \citet{diesing+19,diesing+21} \citep[and references within, in particular][]{malkov97, malkov+00, blasi02, blasi04, amato+05, amato+06, caprioli+09a, caprioli+10b, caprioli12}.  
This framework calculates the CR distribution accelerated at a quasi-parallel, nonrelativistic collisionless shock by concurrently solving for the shock jump conditions and steady-state diffusion--advection equation \citep{skilling75a, bell78a} as a function of time.
It self-consistently considers dynamical modifications induced by both CRs and turbulent magnetic fields via their pressure contributions in the shock jump conditions \citep{caprioli+09a, haggerty+20, diesing+21}.
Within this framework, the maximum proton energy ($E_{\rm max}$) is calculated by imposing a free-escape boundary ahead of the shock at a fixed fraction ($X_0$) of the shock radius \citep{vladimirov+06, caprioli+09b,diesing23} and ensuring that the current in escaping CRs is strong enough to drive the Bell instability to saturation \citep{bell+13, blasi+15}, hence efficiently amplifying the preexisting magnetic field \citep{cristofari+21, cristofari+22}, see \S\ref{subsec:mfa}.

Protons are injected into the acceleration process with momenta above $p_{\rm inj} \equiv \xiinj m_p \vsh \left(1-R^{-1}\right)$, where $\vsh \left(1-R^{-1}\right)$ is the upstream velocity in the downstream frame and $\xiinj$ is a factor taken from kinetic simulations \citep{caprioli+14a, caprioli+15}.  
Hybrid simulations also suggest that ion injection from the thermal pool is significantly reduced in quasi-perpendicular regions, where CRs may be accelerated with relatively steep spectra \citep{orusa+23, orusa+25b}.
Therefore, we adjust proton injection to produce large acceleration efficiencies $\xicr\gtrsim 10\%$ in the polar caps, and much lower $\xicr$ elsewhere, constrained by the $\gamma$-ray emission observed in the NW region \citep{lemoine-goumard+25}.

The kinetic model generates an instantaneous CR proton distribution at each timestep.
The corresponding electron distributions are derived by applying the analytical approach in \cite{zirakashvili+07}, which accounts for radiative losses and requires the normalization of the electron spectrum relative to the proton spectrum ($\Kep$). 
Electron injection into DSA is not completely understood, but typical values of $\Kep$ range from $10^{-2}$ to $10^{-4}$ in SNRs \citep[e.g.,][]{volk+05, slane+14, park+15, sarbadhicary+17}. 
Kinetic simulations of quasi-parallel shocks are consistent with these values \citep[e.g.,][]{park+15, gupta+24b, gupta+25}.
The instantaneous distributions define invariant mass shells of nonthermal particles, which are then evolved to account for adiabatic and synchrotron losses, as detailed by \citet{zirakashvili+07}, \citet{caprioli+10b}, \citet{morlino+12}, and \citet{diesing+19}.
The procedure assumes adiabatic expansion ($PV^{\gamma_{\rm ad}}=const$), with adiabatic index $\gamma_{\rm ad} = 5/3$, and pressure equilibrium in the downstream (such that $P \propto n_{0} \, v_{\rm sh}^2$), causing density to decrease by a factor 
\begin{equation}
    L_{\rm ad} = \left(\frac{n_{0,\rm in} \, v_{\rm sh,in}^2}{n_{0,\rm age} \, v_{\rm sh,age}^2}\right)^{\frac{1}{\gamma_{\rm ad}}}
\label{eq:ad}
\end{equation}
between when a shell was produced ($t_{\rm in}$) and the time considered ($t_{\rm age}$).
We track the evolution of accelerated particles via the time-dependent shells, which (after evolving) we sum to obtain an aggregate multizone spectrum of accelerated nonthermal particles, as in \citet{diesing+19,diesing+21}.




\subsection{Magnetic Field Amplification and Its Effects} \label{subsec:mfa}

Motivated by the observations of fields as large as $100-500$ $\mu$G in young SNRs \citep[e.g.,][]{volk+05,parizot+06,uchiyama+07} and by kinetic simulations of quasi-parallel shocks \citep{reville+13, caprioli+13, caprioli+14a}, we include MFA due to both the resonant streaming instability \citep{kulsrud+68, skilling75a, bell78a, zweibel79, lagage+83a} and the nonresonant instability (or ``Bell instability'') \citep{bell04, amato+09, zacharegkas+24}.  

Given the shock velocity of SN 1006, the Bell instability driven by escaping CRs upstream should be the dominant source of MFA \citep[e.g.,][]{vladimirov+06,caprioli+09a,bell+13, blasi+15, cristofari+20}.
We consider the amplified field at saturation \citep{bell04, zacharegkas+24}, driven by a CR power-law energy distribution $\propto E^{-q}$, with $q>2$, as derived in equation 5 of \citet{cristofari+21}. 
If $\pcr$ is the pressure in CRs at the shock, we can write the upstream magnetic pressure, constant throughout, as
\begin{equation}
 P_{\rm B} = \frac{3}{4}\frac{\vsh}{2c} \left(\frac{E_{\rm max}}{m_pc^2}\right)^{2-q} \frac{\pcr}{(q-1) I(q)} \; ,
\label{eq:PB}
\end{equation}
where numerical factor $I(q)=\int_{1}^{E_{\rm max}/mc^2} x^{2-q}/(1+x^2)^{1/2} dx$ \citep[see equation 2 in][]{cristofari+22}.
We estimate $q \approx 2.2$ from the radio band (see \S\ref{subsec:SED}) and $E_{\rm max} \approx 10^5$ GeV from $\gamma$-ray observations, which together give $I(q = 2.2)\approx4$.

As pointed out by \citet{bell+13} and \citet{cardillo+15}, requiring that the Bell instability driven by the current in escaping CRs saturates in the SNR lifetime places an additional constraint on $E_{\rm max}$, which we satisfy by choosing an appropriate boundary condition in our solution of the Parker equation \citep{caprioli+10b}.
More precisely, we impose that the isotropic part of the CR distribution function vanishes at a distance $X_0 R_{\rm sh}$ ahead of the shock, beyond which CRs around $E_{\rm max}$ escape freely.
We iterated over the value of $X_0$ and found that $X_0=0.02$ produces a maximum energy that adheres to the condition that the age of the SNR be longer than both the time needed to saturate the Bell instability \citep{bell+13, blasi+15,cristofari+22} and the diffusive acceleration time in the amplified magnetic field \citep{lagage+83a, blasi+07}.

Hybrid simulations suggest that downstream magnetic fluctuations, i.e., the CR scattering centers, drift with respect to the thermal plasma at approximately the local Alfv\'en speed ($v_{\rm A}$) away from the shock front \citep{haggerty+20,caprioli+20}.  
In this \emph{postcursor} region, the drift enhances escape from the acceleration region, leading to fluid compression ratios $R>4$ \citep{haggerty+20} and steeper CR spectra \citep[$q>2$,][]{caprioli+20}, consistent with observations of Galactic SNRs \citep[][and references therein]{diesing+21}, including SN 1006 \citep[][]{giuffrida+22}. 
This effect is also present in the near upstream (\emph{precursor}) region, where the drift is again the local Alfv\'en speed moving away from the shock front \citep{zirakashvili+08, caprioli12, diesing+21}, but---since the fluid speed is smaller and magnetic fluctuations are compressed in the downstream---the effect of the precursor is considerably less than that of the postcursor, though nonnegligible.
We include the effects of the postcursor (and precursor) into our kinetic model by adding (subtracting) the local Alfv\'en speed to the downstream (upstream) fluid velocity, $u_2$ ($u_1$), in the diffusion--advection equation \citep{caprioli12, diesing+21}, which modifies the slope, characterized by the total compression ratio $R=\vsh/u_2$, to be
\begin{equation}
 q = \frac{R+2+R(2w_{\rm A,2}-w_{\rm A,1})}{R-1-R(w_{\rm A,2}+w_{\rm A,1})}  > \frac{R+2}{R-1} \;,
\label{eq:slope}
\end{equation}
where $w_{\rm A,i} = v_{\rm A, i}/\vsh$, $v_{\rm A, i} = B_{\rm i}/\sqrt{4 \pi m_{\rm p}n_i}$, and subscripts 1 and 2 denote parameter at near upstream and immediately downstream of the shock, respectively.

Finally, in \S\ref{subsec:radpro_radio} we consider the possibility that the amplified field is damped in the postshock medium \citep{ptuskin+03}, but find a posteriori that no damping is required to account for observations.

\subsection{Particle Diffusion} \label{subsec:pardiff}

At quasi-parallel shocks MFA is expected to produce $\delta B/B_0 \geq 1$, where $\delta B$ is the amplitude of gyroscale magnetic fluctuations, eventually leading to Bohm diffusion, where the pitch-angle scattering rate is comparable to the Larmor frequency at all momenta \citep{reville+13, caprioli+14b, caprioli+14c}. 
However, there is evidence that at quasi-perpendicular shocks Bohm diffusion is not realized \citep[e.g.,][]{rothenflug+04, gamil+08}, leading to a $E_{\rm max}$ generally lower than in the polar caps.

When electron acceleration occurs in the Bohm limit and is limited by synchrotron losses, the X-ray cutoff energy ($E_{\rm cut}$) is independent of the magnetic field strength and $E_{\rm cut} \propto \vsh^2$.
Since the SNR radius in the SE and polar caps is the same, one would conclude that $\vsh$, and hence $E_{\rm cut}$, should be the same, too.  
However, \citet{rothenflug+04}, \citet{gamil+08}, and \citet{li+18} reported a decrease in the X-ray synchrotron cutoff emission from the NE and SW rims toward the SE region. 
Low synchrotron flux and $E_{\rm cut}$, together with the high radio fractional polarization \citep{reynoso+13}, in the quasi-perpendicular regions all point to inefficient MFA \citep[consistent with kinetic simulations, e.g.,][]{caprioli+14a}, and imply $D>D_{\rm B}$ in these regions.
We account for this effect by introducing a ``Bohm factor" $\eta \equiv D/D_{\rm B}\geq 1$, where $D$ is the diffusion coefficient and $D_{\rm B}$ is the Bohm diffusion coefficient \citep{gamil+08}; we generally expect $\eta=1$ in the polar caps and $\eta\gtrsim 1$ elsewhere.

We constrain $\eta$ using the azimuthal dependence of $E_{\rm cut}$ from \citet{miceli+16}, which reported a change in $E_{\rm cut}$ by a factor of about $1.8$ over a $\sim\!20^\circ$ angle (see their figure 3) from the SW quadrant's center to edge.
By extrapolating the relationship between angle and cutoff energy from the SW to the nonpolar caps (over $\sim 90^\circ$) and utilizing the relation $E_{\rm cut} = E_{\rm cut,B}/\eta$, where $E_{\rm cut,B}$ is the cutoff energy with Bohm diffusion (achieved in the polar caps), we estimate that $\eta$ should vary from 1 to about 8.1 throughout the parallel-to-perpendicular regions.
However, it is possible that the diffusion coefficient in the nonpolar caps cannot be described by a simple rescaling of the Bohm diffusion coefficient. 
Hence, we caution that the modeled nonpolar cap spectra are unconstrained outside Fermi-LAT observations \citep{lemoine-goumard+25}.



\subsection{Photon Production}  \label{subsec:phopro}

We create broadband multiwavelength photon spectra from the cumulative particle distributions (see \S\ref{subsec:paracc}) utilizing the Python package for radiative processes \texttt{naima} \citep[][]{naima}, which calculates synchrotron, IC, nonthermal bremsstrahlung, and neutral pion decay emission. 
For IC, we estimate the ambient photon field by combining the contributions of the cosmic microwave background and the infrared+optical background \citep[][]{winkler+13, porter+17}, which sum to 1 $\rm eV \, cm^{-3}$ at the location of SN 1006.

We construct radial maps by integrating the space-dependent emissivity along the line of sight. This entails determining the radial position of each invariant mass shell at $t_{\rm age}$. 
The radial profile is solved by positioning shells assuming no mixing between them, i.e., CR advection dominates over diffusion in the downstream magnetic fields (which are also evolved adiabatically).

Emission from each quadrant is calculated assuming that polar caps subtend cones of solid angle $4\pi f$, where $f$ is a filling factor set based on TeV maps \citep{SN1006HESS,lemoine-goumard+25}, and nonpolar caps subtend cones such that the total solid angle is $4\pi$.

\section{Results} \label{sec:results}

\begin{table}
\begin{ruledtabular}
\begin{tabular}{cccc}
 \multirow{2}{5em}{Parameter} & \multicolumn{3}{c}{Value} \\
 \cmidrule{2-4}
 & NE/SW & SE & NW \\ [0.5ex] \hline 
 & \\[\dimexpr-\normalbaselineskip+1pt]
 $v_{\rm sh,age}$ ($\rm km\,s^{-1}$) & 5300 & 5300 & 2700 \\ [0.3ex]
 $R_{\rm sh,age}$ (pc) & 9.7 & 9.7  & 7.8 \\ [0.3ex] 
 $\nism$ ($\times10^{-2}$ $\text{cm}^{-3}$) & 1.8 & 1.8 & $1.8^{\dagger}$ \\ [0.3ex] 
 \cmidrule{1-4}
 $\xicr$ & $20\%$ & $0.4\%$ & $0.8\%$  \\ [0.3ex]
 $\Btwo$ ($\mu \text{G}$) & 40 & 6 & 10 \\ [0.3ex]
 $R$ & 5.0 & $4.0$ & $4.0$ \\ [0.3ex]
 $\eta$ & 1.0 & 8.1 & 8.1 \\ [0.3ex]
 $f$ & 0.1 & 0.4 & 0.4 \\ [0.3ex]
 $\Kep$ $\left(\times 10^{-4}\right)$ & 9.3 & 9.3 & 9.3 \\ [0.5ex]
\end{tabular}
\end{ruledtabular}
\caption{\label{tab:output} Each quadrant's parameters describing shock evolution (top) and particle acceleration (bottom); 
the NE and SW regions have identical values. 
The acceleration efficiency $\xicr$ is large at the polar caps, which leads to modified shock compression $R > 4$, and less than 1\% elsewhere.
The SED fitting is consistent with having Bohm diffusion ($\eta=1$) in the polar caps, and less effective diffusion in the more oblique regions.
$\Kep$ is held constant.\\
$^{\dagger}$Density jumps to $\ncl = 0.18$ $\text{cm}^{-3}$ at $\Rcl = 5$ pc.
}
\end{table}

\begin{figure*}
    \centering
    \includegraphics[width=1.0\textwidth]{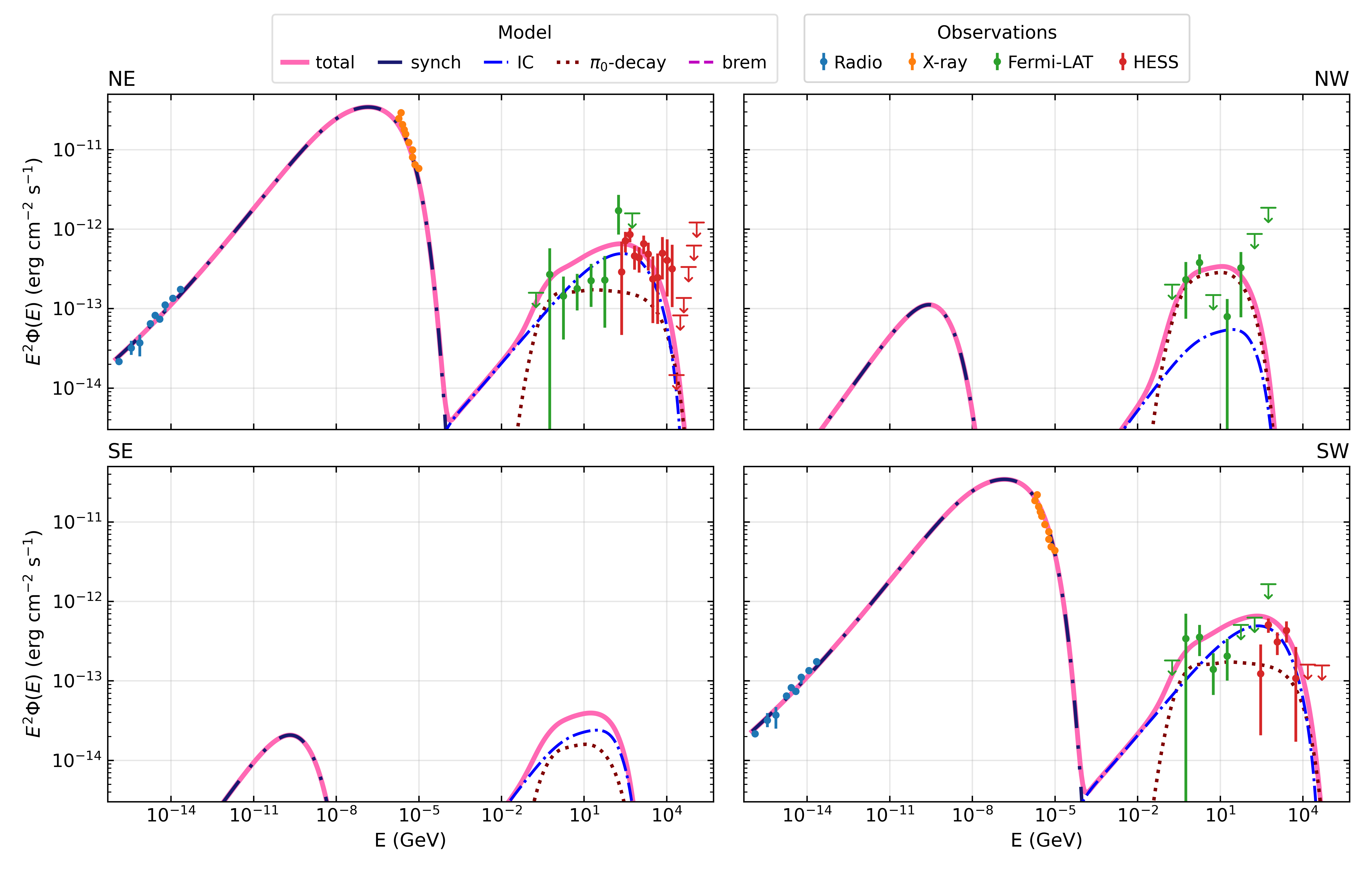}
    \caption{\label{fig:SED} Nonthermal multiwavelength SED for each SN 1006 quadrant (identified in title). 
    Curves correspond to total emission (solid pink), synchrotron (dotted--dotted--dashed dark blue), IC (dotted--dashed bright blue), $\pi^0$-decay (dotted maroon), and bremsstrahlung (dashed magenta, typically negligible).
    The data points show radio \citep[blue, from a selection by][]{allen+01}; X-rays from Suzaku \citep[orange,][]{bamba+08}; $\gamma$-rays from Fermi-LAT \citep[green,][]{lemoine-goumard+25} and HESS \citep[red,][]{SN1006HESS}. 
    We scale the radio and X-ray spectral properties between the polar caps as in \cite{lemoine-goumard+25}, assuming that the FS nonthermal emission is negligible in the nonpolar caps in both radio \citep[e.g.,][]{rothenflug+04,cotton+24} and X-rays \citep[e.g.,][]{koyama+95,long+03,rothenflug+04}.
    Both polar caps are reasonably well fit with the same model, which is essentially leptonic at TeV energies, with a hadronic contribution in the GeV.
    For the nonpolar caps, the modeled nonthermal emission (excluding the NW $\gamma$-rays) represents an upper limit.
    In the dense NW quadrant, a largely hadronic model reasonably fits the observed $\gamma$-ray emission. 
    }
\end{figure*}

\subsection{Photon Spectra} \label{subsec:SED}

We compare the best-fit values (see Table \ref{tab:output}) of the modeled SEDs of each SN 1006 quadrant with the multiwavelength data in Figure \ref{fig:SED}. Our results closely reproduce the observed emission in the polar caps and NW region. 

We find that the radio synchrotron emission ${E^2 \Phi(E) \propto E^{0.41}}$ in the polar caps constrains the proton and uncooled electron spectra to be $\propto E^{-2.18}$, steeper than the DSA test-particle prediction and consistent with the postcursor effect for an amplified magnetic field (see Equation \ref{eq:slope}).
Having inferred the postshock magnetic field to be $B_2 = 40$ $\mu \rm G$ via the radio slope, the normalization of the synchrotron emission constrains the electron/proton ratio $\Kep\sim 9.3\times 10^{-4}$, and in turn the normalization of the bremsstrahlung and IC contributions.
This value of $\Kep$ is in reasonable agreement with the formula inferred from first-principles kinetic simulations \citep{gupta+25}, which for our parameters would be $\sim\!1.1\times 10^{-3}$.


Our results show that the $\gamma$-ray emission from the polar caps is mostly of leptonic origin, with possibly a small hadronic contribution in the GeV band (see Figure \ref{fig:SED}). 
Yet, in the NW quadrant, where the density is a factor of $10$ larger, the Fermi-LAT $\gamma$-rays are likely attributed to $\pi_0$-decay, which restricts the acceleration efficiency in the oblique shock regions to below that of the normal shock regions.  
This provides evidence for a relationship between hadronicity and ambient density (as opposed to acceleration efficiency), consistent with \cite{corso+23}.

Note, the modeled nonpolar cap spectra are not constrained beyond the observed GeV emission \citep{lemoine-goumard+25} (see \S\ref{subsec:pardiff}), though the lack of nonthermal X-rays suggests that the maximum energy does not make it into the TeV range.


Finally, though nonthermal bremsstrahlung is calculated in our model, it is never important for reproducing the observed $\gamma$-ray emission.  
This is not surprising given that bremsstrahlung scales with ambient density, which is very low here. Moreover, even when the ambient density is large, p-p emission typically dominates \citep{aharonian+06,corso+23}.


\subsection{Azimuthal Dependence}  \label{subsubsec:azdep}
Our results corroborate the expected azimuthal dependence of CR efficiency on shock obliquity \citep[e.g.,][]{caprioli+14a, caprioli+15}, with quasi-parallel regions also exhibiting greater MFA via CR-driven magnetic instabilities than quasi-perpendicular regions.  
This conclusion arises from the fact that, in order to fit observations, 
(1) the CR efficiency $\xicr$ is found to be $20\%$ in the polar caps and under 1\% in the nonpolar caps (see Table \ref{tab:output}); 
(2) the steep polar cap spectra ($q=2.18$) return an enhanced compression ratio of $R = 5.0$, whereas the flatter nonpolar cap spectra yield $R=4.0$, consistent with the azimuthal variation in postshock density from X-ray measurements \citep[][]{giuffrida+22};
(3) $B_2$ is amplified in the polar caps, with field strength a factor of $\gtrsim 10$ larger than the interstellar medium (ISM) ($\sim\!3$ $\mu \rm G$), which is sufficient for the precursor region to be comparable to the angular resolution of Chandra and thus undetectable as discussed in \cite{morlino+10}, and minimal in the nonpolar caps;
(4) strongly amplified, turbulent, magnetic fields in the polar caps and merely compressed fields in the nonpolar regions agree with the measured radio fractional polarization levels \citep{reynoso+13}.

The GeV $\gamma$-ray emissions observed in the NW quadrant reveal that, despite being inefficient, particle acceleration must be at work to some degree in quasi-perpendicular regions.
While in principle simple reacceleration of Galactic CRs cannot be ruled out \citep{cardillo+14, caprioli+18}, one promising mechanism is shock drift acceleration, where particles gain energy by gyrating across the shock front repeatedly, which recent 3D kinetic simulations \citep{orusa+23, orusa+25b} have shown to occur spontaneously at quasi-perpendicular shocks for large Mach numbers.

Since the relationship between shock obliquity and CR efficiency is continuous \citep{caprioli+14a,caprioli+18}, we investigate whether accounting for azimuthal variation within a quadrant appreciably affects our results.  
Specifically, we use the analysis presented in \cite{miceli+16} for the synchrotron cutoff energy variation and its link to the local diffusion coefficient to construct obliquity-dependent variations in $\xicr$ and hence in MFA. 
We partition the SW region into two sectors, each with filling factor $f\!=0.05$: a center region with $\xicr = 20\%$, Bohm diffusion, and constant cutoff energy, as well as a pair of edge regions with smaller efficiencies, MFAs, and cutoff energies. 
For Bohm diffusion, the cutoff energy only depends on the shock speed \citep[e.g.,][]{zirakashvili+07}, which must be constant throughout each quadrant. 
Therefore, the $\sim\!1.8$-fold variations reported by \cite{miceli+16} (figure 3 within) suggest the edge regions must have insufficient MFA to produce Bohm diffusion, and $\delta B/B_0 \approx 0.55$.
Within our framework, we tune down injection until the desired $\delta B/B_0$ at $\age$ is achieved, which corresponds to $\xicr = 0.5\%$ in such edge regions.
Having determined each sector's amplified magnetic field and $\xicr$, we increase $\Kep$ to $1.9\times10^{-3}$ to match the normalization of the SW synchrotron emission.
This new value of $\Kep$ is $\sim$2 times larger than the original in Table \ref{tab:output}, but still in agreement with \cite{gupta+25} (within a factor of $\sim\!2$).
All other parameters remain the same.

The resulting SED from the summed contributions of two SW sectors with different CR efficiencies is essentially identical to Figure \ref{fig:SED}.  The only exception is the hadronic component, which decreased as a consequence of the smaller $f$ value of the high-CR-efficiency region, yet still reasonably reproduces Fermi-LAT observations.  
Thus, we determine that accounting for some azimuthal variation within a quadrant does not affect our results significantly.

\subsection{X-ray Radial Profiles}  \label{subsec:radpro_X}

\begin{figure*}
    \centering
    \includegraphics[width=1.0\textwidth]{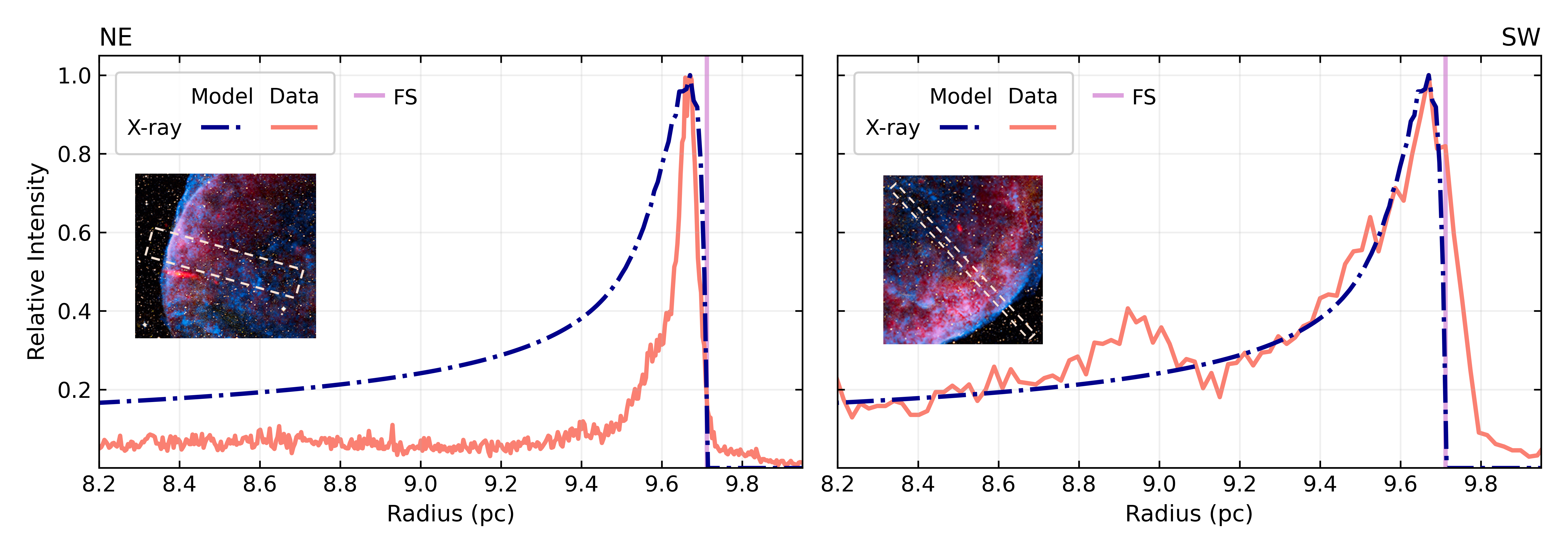}
    \caption{\label{fig:LOS}  X-ray (1-7 keV, dotted--dashed blue) radial profiles for the polar caps (identified in title), with the estimated FS (solid plum) indicated.
    Observed 1-7 keV X-ray profiles (orange, Chandra CDC 641 were extracted from areas indicated by dashed boxes on the inset SN 1006 composite image.
    Each quadrant is scaled individually per photon energy.
    Both quadrants have thin X-ray rims that peak near the FS and are well approximated by our model. 
    }
\end{figure*}

We now investigate the radial profiles of the polar caps in the X-ray (1-7 keV) band.
We do not consider nonpolar cap regions because they are dominated by thermal, as opposed to nonthermal, X-ray emission \citep[e.g.,][]{westerkamp+24}.
Likewise, limited angular resolution thwarts the construction of $\gamma$-ray profiles.

Figure \ref{fig:LOS} shows the polar cap X-ray radial profiles (from a list of Chandra observations contained in the Chandra Data Collection (CDC) 641 \dataset[doi:10.25574/cdc.641]{https://doi.org/10.25574/cdc.641}) taken from the regions in the insets and compared with the model outlined in \S\ref{subsec:phopro}. We acknowledge some uncertainty in selecting profiles due to their azimuthal dependence \citep[][]{westerkamp+24}; however, we consistently extracted profiles in both X-ray and radio (later in \S\ref{subsec:radpro_radio}) from the same regions representative of each polar cap.

The rough agreement suggests the inferred postshock field is sufficiently large to reproduce the thin X-ray rims.
Additionally, we find that changing the amplified field (within spectral constraints) affects the rim width, which reinforces the conclusions of \citet{ressler+14} that rims are determined by synchrotron losses rather than damping. 

It is remarkable that the value of the downstream magnetic field ($B_2\approx 40$ $\mu$G) produced by CR streaming instabilities and further compressed at the shock (\S\ref{subsec:mfa}) accounts simultaneously for four independent observables: the particle slope, the enhanced compression ratio, the thickness of the X-ray rims, and the normalization of the synchrotron emission ($\Kep$ also includes the normalization of the leptonic $\gamma$-ray emission).

\subsection{Radio Radial Profile}  \label{subsec:radpro_radio}

The interpretation of the radio profile (at 1335 MHz), which reflects the entire SNR evolution, instead of only capturing recent activity like the X-ray and $\gamma$-ray profiles, is much more problematic.

\begin{figure*}
    \centering
    \includegraphics[width=1.0\textwidth]{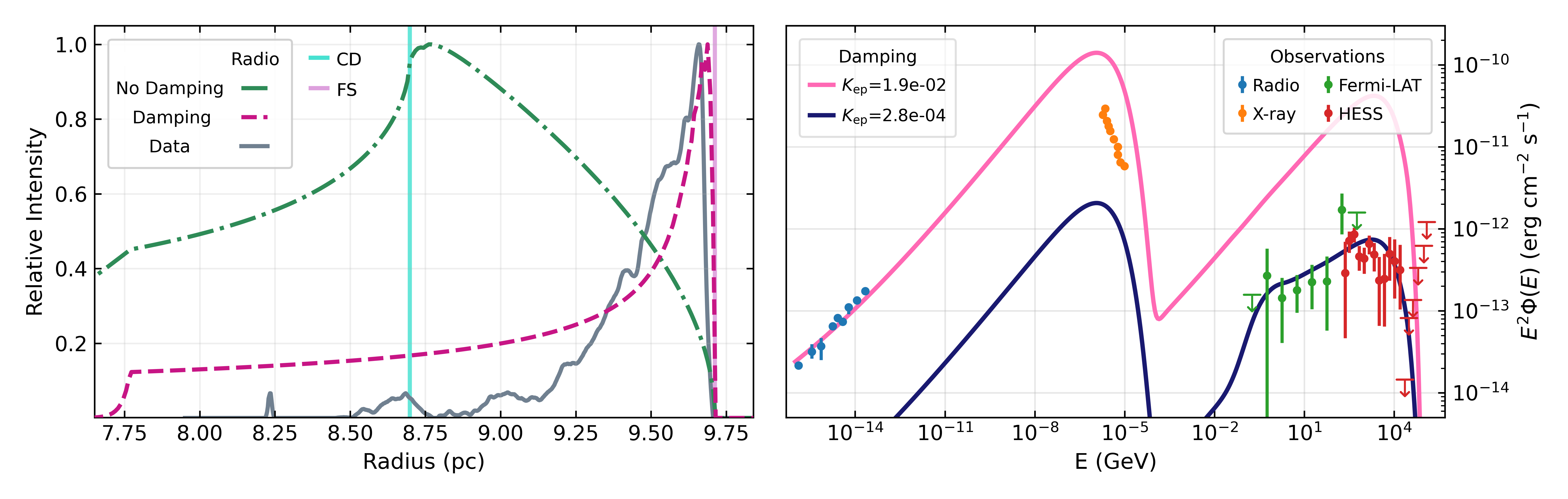}
    \caption{\label{fig:damp}  
    Left panel: NE radio (1335 MHz) profiles for the undamped (dotted--dashed green) and damped model ($\Gamma_{\rm nl} = 2\times10^{-2}$ yr$^{-1}$, dashed magenta).
   The observed profile \citep[solid gray,][]{cotton+24} is extracted from the dashed box in Figure \ref{fig:LOS}.
    The estimated positions of CD (solid cyan) and FS (solid plum) are shown.
    Right panel: fit of the SED (from Figure \ref{fig:SED}) for two damped models corresponding to two different $\Kep$ values (identified in legend, pink and dark blue curves).
    A single $\Kep$ value can account for either the radio or $\gamma$ emission, but not both; hence we rule out damping as the explanation for the width of the radio rims.
    }
\end{figure*}

Figure \ref{fig:damp} (left panel) shows the MeerKat radio profile \citep[][]{cotton+24} compared with our fiducial model (green curve), shifted radially such that the model and observed emission cease at $r >\rsh$, for the NE quadrant. 
As a result of projection effects, our model naturally estimates the radio emission peak at the contact discontinuity (CD). 
In contrast, observations reveal the radio peak to be much closer to the FS, with a width only slightly wider than its X-ray counterpart.


Since synchrotron losses are inconsequential to radio-emitting electrons, one possible explanation for such a thin radio rim may be provided by magnetic field damping \citep[e.g.,][]{sushch+18,wilhelm+20}. 
Similar to \citet{ptuskin+03} and \citet{morlino+12}, we consider nonlinear Landau damping, which is expected to be the dominant form of damping in warm plasmas \citep[e.g.,][]{lee+73, mckenzie+82}.
Though the actual expression for the damping rate $\Gamma_{\rm nl}$ is not fully understood \citep[see, e.g.,][and references therein]{schroer+25a, schroer+25b}, we follow the phenomenological prescription of \citet{ptuskin+03} (equations $10-12$) and write a time-dependent postshock field as $B_2\exp[-\Gamma_{\rm nl}(\age-t)]$.  
A finite $\Gamma_{\rm nl}$ reduces the radio width and shifts the peak position toward the FS.  
We test this solution for the NE quadrant, keeping all other parameters from Table \ref{tab:output} constant, and vary $\Gamma_{\rm nl}$ until the damped model well approximates the observed radial profiles at $\Gamma_{\rm nl} = 2\times10^{-2}$ yr$^{-1}$ (see Figure \ref{fig:damp}).  
Our success in fitting the radio profile, however, comes at the expense of the multiwavelength SED;
the right panel of Figure \ref{fig:damp}) shows how two inconsistent values of $\Kep$, specifically $\Kep=1.9\times10^{-2}$ and $\Kep=2.8\times10^{-4}$, would be needed to jointly fit the radio and $\gamma$-ray emission. 

We conclude from this analysis that we are unable to resolve the discrepancy between modeled radio profiles and observations by simply introducing magnetic field damping.
Instead, we argue that the CD might be much closer to the FS than our model estimates.   
X-ray observations \citep{gamil+08} support this picture, finding the ratio of FS to CD radii to be $R_{\rm FS}/R_{\rm CD}\simeq 1.00$ near the polar caps, and an average of $R_{\rm FS}/R_{\rm CD}\simeq1.04$ in the SE quadrant.  
In comparison, our 1D SNR evolution predicts $R_{\rm FS}/R_{\rm CD}\simeq1.12$ for the polar caps and $R_{\rm FS}/R_{\rm CD}\simeq1.16$ for the SE quadrant; our evolution effectively disregards the vulnerability of the CD to hydrodynamical instabilities, e.g., Rayleigh--Taylor \citep[][and references therein]{blondin-ellison01} and ejecta clumping \citep{rakowski+11,orlando+12}, both of which can drive $R_{\rm FS}/R_{\rm CD}$ closer to unity. 

\section{Conclusion} \label{sec:conclusion}

In summation, we model the multiwavelength spectral and spatial properties of SN 1006 to constrain the relationship between CR acceleration and shock obliquity, and to identify the nature (hadronic or leptonic) of its $\gamma$-ray emission.
Our model divides SN 1006 into four quadrants (Figure \ref{fig:quad}) based on shock inclination and ambient density profile, with the NW region having a larger density because of a recent collision with a dense cloud \citep[][]{acero+07}.
The shock evolution in each quadrant is constrained using SN 1006's morphology (see \S\ref{subsec:hydro}).

We use a semianalytic kinetic model of particle acceleration at nonrelativistic shocks to produce multizone CR spectra \citep{caprioli+09a, caprioli+10b, caprioli12, diesing+19, diesing+21}, which are evolved considering adiabatic and synchrotron losses.
Our framework includes nonlinear modifications to DSA to account for efficient CR acceleration and CR-driven MFA.
It also incorporates the postcursor effect \citep[from][]{haggerty+20,caprioli+20}, which is essential for rendering the steep spectra \citep[][]{allen+01, lemoine-goumard+25} and shock compression ratios \citep{giuffrida+22} observed.
From the CR distributions, we construct multiwavelength photon spectra, see Figure \ref{fig:SED}, and radial profiles, see Figures \ref{fig:LOS} and \ref{fig:damp}, using the model parameter values in Table \ref{tab:output}.  
We conclude the following:

\begin{itemize}
    \item The observed multiwavelength spectral and X-ray spatial properties of SN 1006 can be reproduced with a self-consistent model of particle acceleration based on kinetic simulations with fewer free parameters than observable quantities.
    This agreement supports strong MFA in the polar caps via the Bell instability driven by escaping CRs upstream and drifting magnetic fluctuations with respect to the background plasma (the postcursor effect).

    \item The downstream magnetic field of $B_2\approx 40$ $\mu$G in the polar caps simultaneously accounts for the CR spectral slope ($q\sim 2.2$), modified shock hydrodynamics ($R\sim 5$), X-ray rim widths, and  synchrotron normalization.
    
    \item The nonthermal emission from SN 1006 reflects the relationship between shock obliquity and CR acceleration efficiency inferred from kinetic simulations \citep[e.g.,][]{caprioli+14a}.
    The SEDs of different quadrants are accounted for with efficient CR acceleration ($\xi_{\rm CR} = 20\%$) and MFA in quasi-parallel regions, and inefficient CR acceleration ($\xi_{\rm CR}<1\%$) and negligible MFA in quasi-perpendicular regions, consistent with the azimuthal variation in radio fractional polarization \citep{reynoso+13}.

    \item The majority of the $\gamma$-ray emission from SN 1006 is leptonic, although hadronic emission dominates in the NW region where the ambient density is enhanced---even though $\xi_{\rm CR}$ is small---attesting to the importance of the environment in determining hadronicity  \citep{corso+23}.
    
    \item While the narrow X-ray rims support synchrotron losses dictating rim width, the thin radio rims require another explanation.
    We rule out the possibility that such a width is controlled by magnetic field damping because it is inconsistent with the observed SED (Figure \ref{fig:damp}).
    A compelling alternative explanation for the observed radio profile is that the CD is prone to Rayleigh Taylor instabilities and/or ejecta clumping \citep{gamil+08,miceli+09}, which decreases the distance between the FS and CD.
\end{itemize}

\begin{acknowledgements}
 This work was supported in part by NASA grant 80NSSC18K1726, and NSF grants AST-2510951 and AST-2308021. We thank the referee and editors for thoughtful comments. 
\end{acknowledgements}

\bibliographystyle{aasjournal}
\bibliography{Total}


\end{document}